\documentclass[12pt,preprint]{aastex}

\shorttitle{CO evolution and $^{13}$C yield in Nova Oph 2017 }
\shortauthors{Joshi, Banerjee \& Srivastava}
\begin{document}
\title{Nova Ophiuchus 2017 as a probe of $^{13}$C nucleosynthesis and carbon monoxide formation and destruction in classical novae
}

\author{Vishal Joshi \altaffilmark{1}}
\altaffiltext{1}{Physical Research Laboratory, Navrangpura, Ahmedabad, Gujarat 380009, India}
\author{D.P.K. Banerjee \altaffilmark{1}}
\author{Mudit Srivastava \altaffilmark{1}}

\email{vjoshi@prl.res.in, orion@prl.res.in, mudit@prl.res.in}

\begin{abstract}
We present a series of near-infrared spectra of Nova Ophiuchus 2017 in the $K$ band that record the evolution of the first overtone CO emission in unprecedented detail. Starting from  11.7d after maximum, when CO is first detected at great strength, the spectra track the CO emission to +25.6d by which time it is found to have rapidly declined in strength by almost a factor of $\sim$ 35. The cause for the rapid destruction of CO is examined in the framework of different mechanisms for CO destruction viz. an increase in photoionizating flux, chemical pathways of destruction or destruction by energetic non-thermal particles created in shocks. From LTE modelling of the CO emission, the $^{12}$C/$^{13}$C ratio is determined to be 1.6 $\pm$ 0.3. This is consistent with the expected value of this parameter from nucleosynthesis theory for a nova eruption occuring on a low mass ($\sim$ 0.6 M$_\odot$) carbon-oxygen core white dwarf. The present $^{12}$C/$^{13}$C estimate constitutes one of the most secure estimates of this ratio in a classical nova.
\end{abstract}


\section{Introduction}

We present a series of near-infrared (NIR) spectra recording carbon monoxide (CO) emission from the classical nova (CN) Nova Ophiuchus 2017 which are analysed with a two-fold motivation. The first is to make a robust estimate of the $^{13}$C yield and compare it with expected values from theoretical nucleosynthesis models. The second aim is to record the evolution of the CO emission in the nova and thus advance our understanding of the formation and destruction processes of CO in nova winds. One of the striking predictions of nucleosynthesis theory of nova physics is that novae contribute almost all (or all) of the $^{13}$C found in the Galaxy. An additional sensational part of these predictions concerns the extent of the $^{13}$C enrichment viz. the $^{13}$C generation in novae ejecta can be so extreme that the $^{12}$C/$^{13}$C ratio can be smaller than unity (the solar value is $\sim$ 90). The predicted novae value of the $^{12}$C/$^{13}$C ratio are however model dependent on the white dwarf (WD) mass, its core composition (whether it is composed of a carbon-oxygen or ONe core) and the extent of mixing of the accreted envelope with the white dwarf surface material.
The overabundance of $^{13}$C in novae is a consequence of the following. The synthesis of $^{13}$C commences once the thermonuclear runaway (TNR) is underway through the CNO cycle reaction $^{12}$C (p, $\gamma$)$^{13}$N. It hence follows that an initial higher content of $^{12}$C would favor the synthesis of larger amounts of $^{13}$C. Thus $^{13}$C enhancements are favored in  CNe with carbon-oxygen core WDs when in addition there is substantial mixing between nova material and accreted material (henceforth CO stands for  carbon monoxide). The subsequent evolution of the $^{13}$C production and destruction are through $^{13}$N($\beta$$^{+}$)$^{13}$C and $^{13}$C (p,$\gamma$)$^{14}$N respectively. The $\beta$$^{+}$ unstable nuclei $^{13}$N, in the former reaction, is one of the  most overabundant species produced at the peak temperatures of the TNR (Starrfield et al. 1972). Overproduction factors, relative to solar, in the models computed by Jose \& Hernanz (1998) lie in the range 900-2500 for the different carbon-oxygen core models and in the range 400-900 for the ONe models.

To confirm the predictions for the $^{13}$C yield in CNe, direct measurements may be done in two ways. The first is through isotopic analysis of pre-solar carbon-rich grains from meteoritic samples after establishing from other isotopic signatures that the grains under consideration have indeed originated in a nova outburst and not, for example, from an AGB star or a supernova. A second, more direct method, is through modeling the CO emission that a few novae exhibit early after their eruption. Modeling has so far been restricted only for the first overtone $\Delta$$\nu$ = 2 bands which lie between 2.29 - 2.5 $\mu$m. The fundamental band at 4.67 $\mu$m may also be used, but it is observationally challenging because of the strong thermal background in that region. A table is presented later which lists the predicted $^{12}$C/$^{13}$C values from theoretical different models juxtaposed with observed values to enable a comparison between the two.

The second aspect we study is the formation and evolution of CO in the nova ejecta. There is generally a lack of observational data related to molecule formation in nova outflows. The first molecular species to be detected in the ejecta of a nova was CN, as seen in DQ Her, (Wilson $\&$ Merrill 1935). At much later epochs,  H$_2$ (2.122 $\mu$m) was detected in the DQ Her remnant (Evans 1991). Subsequently, with access to the mid-IR/FIR made possible through the advent of space observatories like Spitzer, many  UIR bands have been observed and studied in detail (Evans \& Gehrz 2012). CO emission in novae, specifically, is an extremely transient phenomenon; once it is formed it is rapidly destroyed. The typical time scale of the emission ranges from a few days to around 2-3 weeks, making it easy to escape detection. CO first overtone detections in novae are not too common - there are only 10 reported in all. Even more infrequent are multi-epoch observations during the emission phase. Of these, even less frequent are those with sufficient cadence to document the formation and destruction of the CO emission in detail. These last data having a good sampling are specially vital to enable a comparison with theoretical studies which predict the time evolution of the CO emission. In the present work we present a series of spectra, on a nearly daily cadence, that show the striking evolution of the CO emission in Nova Oph 2017. LTE modelling of the data is done to estimate the physical parameters of the CO gas viz temperature, mass, velocity and the $^{12}$C/$^{13}$C ratio. We also pinpoint the likely mechanism that is likely responsible for the rapid destruction of the CO emission.


\section{Observations:}
 NIR spectra in the 0.85--2.45 $\mu$m region were obtained a at a resolution of $\sim$ 1000 using the Near Infrared Camera Spectrograph (NICS) deployed on the 1.2m telescope of the Mount Abu Observatory, India. Since the observational procedures related to spectroscopy with NICS have been described in detail in several places ( Banerjee et. al. 2014; Joshi et. al. 2015;  Srivastava et al. 2016) we refer the reader to these works. Reduction and analysis of the spectra were done using a combination of IRAF and Python routines developed by us. In the present study, we use only the $K$ band spectra -- the remaining $J$ and $H$ band spectroscopic data and $JHK_{s}$ photometry will be presented elsewhere. The log of the observations is given in Table 1.

\section{Nova Ophuchii 2017}
Nova Oph was discovered by K. Itagaki ( CBAT "Transient Object Followup Reports at http://www.cbat.eps.harvard.edu) on 2017 May 08.7511 and spectroscopically confirmed to be a FeII nova by Williams \& Darnley (2017) on 2017 May 11.15 UT (ATel 10366). Strader et al. (2017) pointed out that pre-discovery ASAS-SN records show that the nova was first detected at V=14.9 on Apr 21.43 UT, seventeen days before the discovery by K. Itagaki. The subsequent ASAS-SN light curve after Apr 21.43 shows substantial variability between V$\sim$ 14 and V$\sim$ 16 with maximum being reached on Apr 30.24 at V = 14.1. But Strader et al (2017) add the caveat that the peak is poorly defined because of large optical variability around maximum which resembles the behaviour seen in e.g. the gamma ray-detected nova V1369 Cen. We take April 30.14 as the reference point for the origin of time. Three days after discovery, NIR spectra were reported independently confirming the FeII class and also reporting first overtone CO emission from the nova (Joshi $\&$ Banerjee, 2017). This was followed by a report on the commencement of dust formation around mid June 2017 ( Joshi, Banerjee $\&$ Srivastava, 2017).

\section{Results}
\subsection{The $^{12}$C/$^{13}$C ratio in Nova Oph 2017}
The top panel of Figure 1 shows a representative spectrum of the novae between 0.8 to 2.5 $\mu$m. The spectrum is very typical of a FeII class of nova (or equivalently Carbon-Oxygen core nova), several spectra of which are shown in Banerjee \& Ashok (2012). The lower panels of Figure 1 shows just the $K$ band displaying the rise and fall of the CO emission along with model fits whose parameters are given in Table 2. Model fits to the CO emission have been obtained assuming LTE populations for the ro-vibrational states and also assuming that the CO emission is optically thin. All the ro-vibration transitions of the first overtone band, up to $\nu$ = 20 and J = 149 are considered with Einstein A coefficients for these transitions being taken from Goorvitch (1994). The model has been applied to several novae viz. V2615 Oph (Das, Banerjee $\&$ Ashok 2009), V496 Sct (Raj et al. 2012), V5584 Sgr (Raj et al. 2015) and V5668 Sgr (Banerjee et al. 2016). Greater details on the model are given in Das et al. (2009). Only  $^{12}$C$^{16}$O and $^{13}$C$^{16}$O are considered as contributing species to the CO emission; the contribution of other isotopologues, such as $^{12}$C$^{17}$O, $^{14}$C$^{16}$O, is assumed to be negligible based on expected abundances of $^{17}$O, $^{14}$C etc in nova ejecta (Jose \& Hernanz 1998).

Figure 2 shows the effect of varying the $^{12}$C/$^{13}$C ratio on the fits to the CO emission. Although this illustration is shown for the spectrum of 12.9 May, similar results are obtained on the other days. The effect of varying the $^{13}$C content is felt most in the 4-2 and 5-3 regions covering the third and forth  $^{12}$CO bands where first two (2-0 and 3-1) bands of  $^{13}$CO starts contributing. In further redward bands, the $^{13}$C effects are not as clearly seen because longward of this, the S/N of the spectra sharply drops because of strong increasing telluric residuals, and shortward of this $^{13}$CO has no emission. The fit to the first of the  $^{12}$CO bands i.e the 2-0 band with its band head at 2.2935 $\mu$m, is undermined by the presence of a neutral carbon line at 2.2906 $\mu$m. This CI 2.2906 $\mu$m line is routinely seen in the spectra of FeII novae at similar strength (amplitude) as the CI 2.1023 $\mu$m line or at lesser strength than the blend of CI lines between 2.1156-2.1295  $\mu$m (Evans et al. 1996, Banerjee \& Ashok 2012 and references therein; Srivastava et al 2015), both of which are marked in Figure 1. A strong CI 2.2906 $\mu$m feature seriously affects the fit of the 2-0 band as seen in the cases of V2274 Cyg (Rudy et al. 2003), V705 Cas (Evans et al. 1996 ), and V2165 Oph (Das et al. 2009). Fortunately, in this nova the CI 2.2906 $\mu$m line is weak and hence its effect on the 2-0 band head is small. By allowing the temperature, velocity and $^{12}$C/$^{13}$C ratio to vary and using a chi-squares minimization criterion for the goodness of the fit, we find that the best formal fit is obtained for a $^{12}$C/$^{13}$C ratio of 1.6 $\pm$ 0.3. Even visual examination clearly establishes how the quality of the fits are sensitive to small changes in the $^{12}$C/$^{13}$C ratio. This is one of the most secure estimates of the $^{12}$C/$^{13}$C ratio with a few positives over some of the earlier estimates. These factors include the fact that (i) contamination of the CO emission with other atomic lines (e.g CI 2.2906 $\mu$m) is minimal (ii) the dependence of the model fits on a varying $^{12}$C/$^{13}$C ratio is demonstrated clearly here in Figure 2 which we believe, has not been done in earlier studies and (iii) the data were obtained at higher spectral resolution compared to some of the other studies (e.g. R = 300 for the V2274 Cyg spectrum in Rudy et al. 2003). We compare our $^{12}$C/$^{13}$C estimate with theoretical predictions for Carbon-Oxygen novae in Table 2. The comparison indicates that the $^{12}$C/$^{13}$C ratio is consistent with the low values predicted from nucleosynthesis theory and specifically suggests that the nova eruption took place on a low mass white dwarf with M(WD) close to = 0.6 Msun. Higher WD masses are not supported in this nova, and for that matter, in none of the other novae in Table 2. We find this thought provoking and perhaps even unusual that all $^{12}$C/$^{13}$C determinations made so far have
never shown a nova with a $^{12}$C/$^{13}$C ratio less than unity which is routinely expected from nucleosynthesis models. Does this mean that all CO producing novae have small masses ( $\sim$ 0.6Msun) or is there some lacuna in the nucleosynthesis predictions. This issue needs the attention of theorists.

 In this context it should be added that Haenecour et al (2016) have recently reported the in situ identification
of two unique presolar graphite grains from the primitive meteorite LaPaz Icefield 031117. One of these grains (LAP-
149) is extremely $^{13}$C-rich and $^{15}$N-poor with a $^{12}$C/$^{13}$C ratio of 1.41 $\pm$ 0.01 which is one of the lowest values
ever observed in a presolar grain. Although such low $^{12}$C/$^{13}$C ratios can be produced in a few astronomical sources viz. born again
AGB stars, J-type carbon stars, novae and core-collapse Supernovae of Type II, Haenecour et al (2016) rule out an origin
in these other sources based on other isotopic signatures. The isotopic compositions of LAP-149 best match an
 origin in the ejecta of a low-mass CO nova. In particular there is a very close match between the $^{12}$C/$^{13}$C = 1.41 $\pm$ 0.01
 and  $^{14}$N/$^{15}$N = 941 $\pm$ 81 values with those from nucleosynthesis predictions for a 0.6Msun Carbon-Oxygen core WD
 predicted to have $^{12}$C/$^{13}$C = 2
 and  $^{14}$N/$^{15}$N = 979 respectively. They thus conclude that grain LAP-149 is the first putative nova grain that quantitatively
best matches nova model predictions, providing the first strong evidence for graphite condensation in nova ejecta.

\subsection{Evolution of the CO emission}
There are less than a handful of novae where the evolution of the CO has been witnessed in
sufficient detail. V705 Cas presented an unique case wherein strong CO was seen in emission just 6d after discovery (and 1d before maximum!)
making it the earliest CO detection. By day 26.5 the CO emission had waned and by day 45 it was below detection.
Raj et al. (2012) documented the CO evolution in V496 Sct between +15d to +21d and saw a drastic drop in
the emission strength during that time. The best sampled evolutionary history was recorded in the case of V2615 Oph (Das et al. 2009) who
found the CO strongly in emission at +5d after maximum. Subsequently the CO emission
remained in a saturated phase for a period of 7 days followed by a rapid decline in strength. Within a month of its first detection the
emission had faded below detection limits. The collective behavior of all the nova described above
are in good agreement with the theoretical predictions of the Pontefract and Rawlings (2004; henceforth PR2004) model discussed below for the evolution of CO.

As summarized in Das et al (2009), the early theoretical studies of the chemistry of novae outflows
were done by Rawlings (1988) in the form of pseudo-equilibrium chemical
models of the pre-dust-formation epoch. These models needed that the outer parts of the ejecta
be dense and carbon be neutral for substantial molecule formation to occur.
In a neutral carbon region, the carbon ionization continuum, which extends to less than 1102 \AA, shields
several molecular species against the dissociative UV flux from
the central star. A modified version was subsequently presented by PR2004 which yielded a result that was a major point of departure from
their earlier model. They found that contrary to the findings of their previous studies,
the formation, evolution and abundances of various molecular species was essentially not photon-dominated
but rather controlled by neutral-neutral, ion-molecule and other chemical reactions. For example, the initial
 primary loss routes for CO were through reactions with H and O+ while at later times the primary loss channels were
 reactions with N+. We show that this might apply in the case of Nova Oph 2017.
 A further significant result in PR2004 is the prediction of the evolution of
the fractional CO abundance with time. It is seen that in both their C or O rich models, the CO abundance remains
constant up to about 2 weeks after outburst i.e. the CO is saturated with
 all the available oxygen or carbon, whichever
has the lower abundance, being completely used up into forming
CO. Following this there is a sharp decline in the CO abundance by a factor of 1000 in $\sim$ 27 d for Model A and a decrease by
a factor of 100 in $\sim$ 16 d for Model B. Essentially a very rapid destruction of CO is predicted
which is observationally confirmed in the novae discussed above (V496 SCt, V705 Cas and V2615 Oph).

Figure 3 shows the decline in the CO flux with time. As a proxy for the CO flux, we have measured the flux under the curve between 2.29 to 2.403 $\mu$m
(i.e. regions including the 2-0,3-1,4-2 and 5-3 bands but excluding the noisy regions further redward).
The rapid decline in the CO strength by a factor of $\sim$ 33 in 14 days is quite remarkable entrenching further
the fact that CO emission in novae is a very transitory phenomenon. In contrast, the strengths of Br$\gamma$ and the NaI 2.2056, 2.2084 $\mu$m
 lines have remained
fairly constant. The behavior of the latter lines is revealing about the possible cause for the rapid destruction of CO. Sodium
 has a first ionization potential (IP) of 5.139 eV; the lowest among those elements whose lines are seen in the NIR spectrum of novae (a comprehensive list of these lines is given in Das et al. 2008). It is easy to show from LTE calculations that
 at around 2500 K, only about 50 percent Na remains
neutral; by 3000 K almost 99 percent is ionised. This automatically implies that strong neutral NaI emission
must originate from relatively cool zones, close to the temperature
of $\sim$ 1800 K where the first dust condensates (carbon)
are expected to form. These Na lines were thus proposed (Das et al. 2008) to be harbingers of imminent dust production in novae
and observations of several dust forming novae have established this to be true. CO has a dissociation energy D(CO) = 11.1 eV, much larger than the first IP
of NaI. It is therefore unlikely that the destruction of CO was caused by an increase in the UV photon flux from the central WD remnant; such an increase would have
destroyed the NaI line emission too. There is supporting evidence that the temperature of the central source hardly changed during the time of the CO destruction. From the AAVSO database and also the SMARTS observations (please add a footnote http://www.astro.sunysb.edu/fwalter/SMARTS) the V band magnitudes between 10 May to 25 May were almost constant at $\sim$ V = 15.1 $\pm$ 0.15. This would imply very marginal changes in the WD temperature using the Bath \& Shaviv, 1976 empirical relation $T$ = 15280$\times$10$^{\Delta m/7.5}$ K between the change in magnitude $\Delta$$m$ and the WD temperature.

We explore whether CO could be destroyed by high energy non-thermal particles from shocks as has been proposed by Derdzinski, Metzger $\&$ Lazzati (2016). In their recent work on dust and CO formation in nova outflows, they
have proposed that dust formation could occur within the cool, dense shell formed behind shocks in a nova wind.
The high densities (n$_{e}$ $\sim$ 10$^{14}$ cm$^{-3}$ ) due to radiative shock compression
result in CO saturation and rapid dust nucleation.
The detection of several gamma-ray emitting novae in the last 5 years has led to mounting evidence for the presence of extremely strong shocks in nova winds as parcels of outflowing matter collide with each other (Ackermann et al. 2014). Acceleration across these shocks can lead to
particles with high energies (even upto the TeV range if $\gamma$ ray emission is to be explained). Derdzinski et al. (2016) propose that CO could be destroyed by such accelerated non-thermal particles. However, we believe that such non-thermal
particles would be even more likely to ionize Na and destroy the neutral NaI emission that
is seen to persist even as the CO emission declines sharply. If we eliminate photoionization and shocks as the
routes for CO destruction in Nova Oph 2017, then the surviving mechanism is through chemical pathways
as suggested by PR2004. We may thus summarize that the fairly unique data presented here, could open new avenues for modelling the evolution of CO in CNe.

Research at the Physical Research Laboratory, India is supported by the Department of Space, Government of India.


\newpage
\begin{figure*}
\plotone{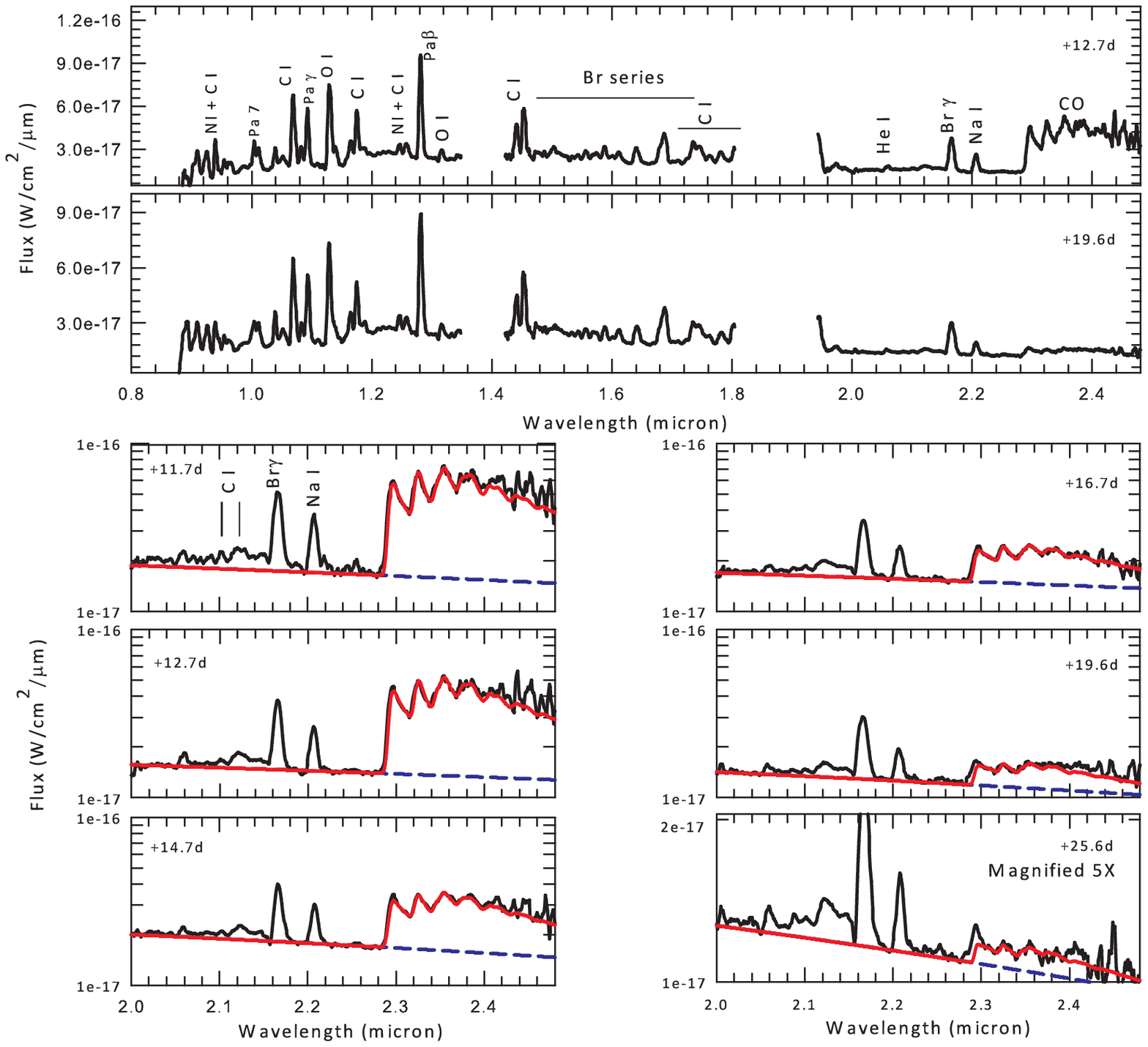}
 \label{fig1}
\end{figure*}

\newpage
\begin{figure}
\epsscale{0.5}
\plotone{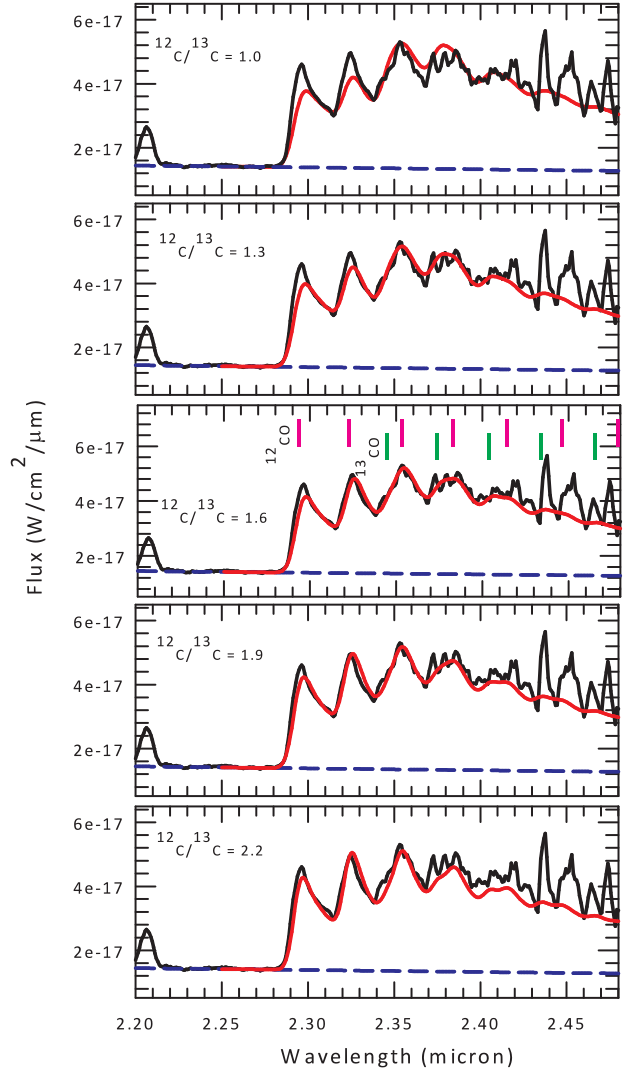}
\caption{Change in quality of the model fits with the $^{12}$C/$^{13}$C ratio. The best fit is
obtained for $^{12}$C/$^{13}$C = 1.6. The position of the $^{12}$CO and $^{13}$CO bandheads are marked.}
 \label{fig1}
\end{figure}

\newpage
\begin{figure}
\epsscale{0.7}
\plotone{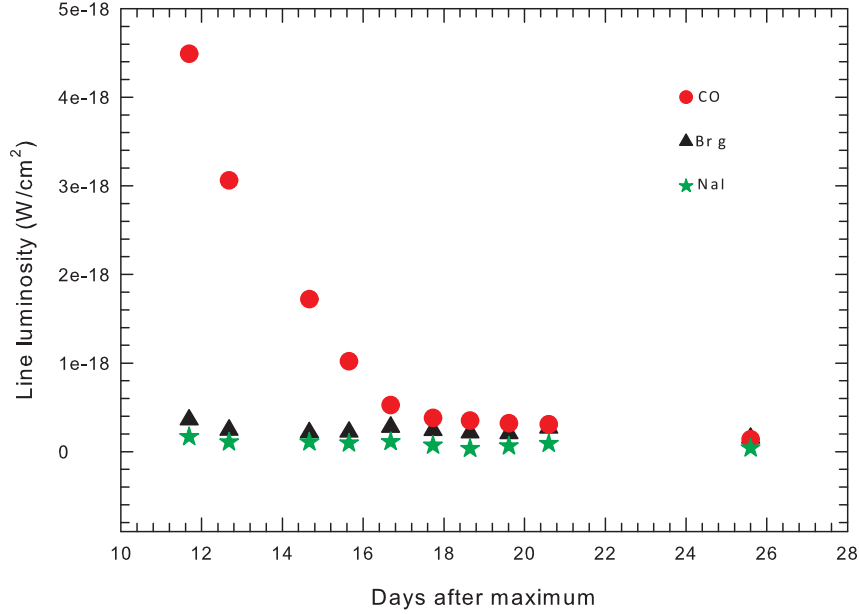}
\caption{Evolution of the  CO flux (measured between 2.29 and 2.403 $\mu$m i.e covering the 2-0,3-1,4-2 and 5-3 bands) with time.
Also shown is the evolution of the Br$\gamma$ and NaI 2.2056, 2.2084 $\mu$m lines.}
\end{figure}

\newpage
\begin{table}
\caption{Log of the  observations}
\centering
\begin{tabular}{c c c c c c c}
\hline
Date & JD& Days after & Int.&Airmass &  Airmass \\
& &maximum & time(s)&Nova &  Standard \\
\hline
2017 May 11.93 	&	2457885.43	&	11.69	&	380	&	1.60	&	1.54	\\
2017 May 12.92 	&	2457886.42	&	12.68	&	2280	&	1.57	&	1.52	\\
2017 May 14.91 	&	2457888.41	&	14.67	&	1520	&	1.57	&	1.53	\\
2017 May 15.89 	&	2457889.39	&	15.65	&	950	&	1.54	&	1.51	\\
2017 May 16.92 	&	2457890.42	&	16.68	&	1520	&	1.59	&	1.53	\\
2017 May 17.97 	&	2457891.47	&	17.73	&	2090	&	1.92	&	1.78	\\
2017 May 18.89 	&	2457892.39	&	18.65	&	1900	&	1.55	&	1.53	\\
2017 May 19.85 	&	2457893.35	&	19.61	&	1900	&	1.56	&	1.62	\\
2017 May 20.84 	&	2457894.34	&	20.60	&	2280	&	1.57	&	1.66	\\
2017 May 25.84 	&	2457898.46	&	25.60	&	3800	&	1.57	&	1.54	\\

\hline
\tablenotetext{a}{Standard star used was SAO 186061 (A0 type).
Maximum was on April 30.24 = JD 2457873.74}
\end{tabular}
\end{table}

\newpage

\scriptsize
\rotate
\begin{table*}
\rotate
\scriptsize
\caption[]{CO emission parameters in NOph 2017;  details of all $\Delta$$\nu$=2 CO detections; predicted model $^{12}$C/$^{13}$C values.}
\begin{tabular}{lrllrcllrclcll}
\hline
 CO Parameters$^a$ && $|$&&& $\Delta$v=2 CO &&$|$&& Predicted $^{12}$C/$^{13}$C &&&&\\
Nova Oph 2017 &&$|$&&&detections&&$|$&&C-O WD models &&&&\\
\hline
Date (days$^b$ &   M(CO)$^c$&$|$ &Nova & Detection   &Observed  & Reference&$|$&$M_{WD}$ &Mixing &&$^{12}$C/$^{13}$C& \\
after $V_{max}$ ) &  (M$_\odot$)&$|$&& epoch$^d$ (d)&$^{12}$C/$^{13}$C&&$|$&(M$_\odot$)&Fraction &H$^e$&JH&S\\
\hline
11.93 May (+11.7d) &1.4e-09&$|$&NQ Vul               &         19& $\geq$3    & Ferland 1979&$|$&0.6 &0.5 &2&-&2.38\\
12.92 May (+12.7d) &1.1e-09&$|$&V842 Cen          &         25&$\sim$2.9    &  Wichmann 1991 &$|$&0.8 &0.25 &0.4 & 0.41&-\\
14.91 May (+14.7d) &6.7e-10&$|$&V705 Cas          &           6&$\geq5$    &  Evans 1996 &$|$&0.8 &0.5 &0.6 & 0.48&1.22\\
15.89 May (+15.7d) &5.0e-10&$|$&V2274 Cyg        &         17&$\sim$1.2    & Rudy 2003   &$|$&1 &0.25 &0.5 &-&-      \\
16.92 May (+16.7d) &4.5e-10&$|$&V2615 Oph        &          9&$\geq$2   & Das 2009 &$|$&1 &0.5 &0.5 & 0.28&0.42\\
17.97 May (+17.7d) &3.80e-10&$|$&V5584 Sgr         &        12  &   & Raj 2014 &$|$&1.15 &0.25 &0.9 & 0.66&-\\
18.89 May (+18.7d) &2.9e-10&$|$&V496 Sct           &        19&$\geq$1.5    & Raj 2012; Rudy 2009 &$|$&1.15 &0.5 &0.7 & 0.50&-\\
19.85 May (+19.6d) &2.6e-10&$|$&V2676 Oph       &        37     & & Rudy 2012a &$|$&1.15 &0.75 & - & 0.36 &-  \\
20.84 May (+20.6d) &2.1e-10&$|$&V1724 Aql       &          7      &&  Rudy 2012b&$|$&1.25&0.5&-&-&0.84\\
25.84 May (+25.6d) &7.1e-11&$|$&V5668 Sgr       &          12    & $\sim$1.5& Banerjee 2016&$|$ \\
 &&$|$&N Oph 2017       &          12   & 1.6& This work&$|$ \\

\hline
\hline
\end{tabular}
\label{table4}
\begin{list}{}{}
 \item (a) On all days the temperature and velocity of the CO gas were found to be very similar and in the range  2400 $\pm$ 200 K and
 1000 $\pm$ 100 km/s respectively (b) : $V_{max}$ on 2017 April 30.14    (c) : For distance $d$ =  1 Kpc. For any other distance, mass scales as $d^{2}$. (d) : In terms of days after discovery.
    (e) :  H = Haenecour et al. 2016; JH= Jose and Hernanz (1998); S = Starrfield et al. 1997
\end{list}
\end{table*}


\begin{thebibliography}{10}

\bibitem[]{}Ackermann, M., et al, 2014, Science, 345, 554
 

\bibitem[]{b4}Banerjee D. P. K., Ashok N. M., 2012, BASI, 40, 243


\bibitem[]{b5}Banerjee D. P. K., Joshi V., Venkataraman V., Ashok N. M., Marion G. H., Hsiao E. Y., Raj A.,2014, ApJL, 785, L11


\bibitem[]{b5}Banerjee, D. P. K.; Srivastava, Mudit K.; Ashok, N. M.; Venkataraman, V., 2016, MNRAS, 455, L109

\bibitem[]{}Bath, G. T., Shaviv, G., 1976, MNRAS, 175, 305
\bibitem[]{}Das, R.K, Banerjee D. P. K., Ashok N. M., 2006, ApJ,653, L141

\bibitem[Das et al.(2008)Das et al.]{2008MNRAS...391...1874} Das R.~K., Banerjee D.~P.~K., Ashok N.~M.,Chesneau O., 2008, MNRAS, 391, 1874

\bibitem[Das et al.(2009)Das et al.]{2009MNRAS...398...375}
 Das R.~K., Banerjee D.~P.~K., Ashok N.~M., 2009, MNRAS, 398, 375

\bibitem[]{b5} Evans A., 1991, MNRAS, 251, 54P

\bibitem[Evans et al.(1996)Evans et al.]{1996MNRAS...282...1049}
 Evans A., Geballe T.~R., Rawlings J.~M.~C., Scott A.~D., 1996, MNRAS, 282, 1049  

\bibitem[]{b5}Evans, A.; Gehrz, R. D., 2012, BASI, 40, 213


\bibitem[Ferland at el.(1979)Ferland et al.]{1979ApJ...227...489}
 Ferland G.~J., Lambert D.~L., Netzer H., Hall D.~N.~B., Ridgway S.~T., 1979, ApJ, 227, 489

\bibitem[]{b5} Goorvitch D., 1994, ApJS, 95, 535

\bibitem[]{b5}Haenecour, H. et al., 2016, ApJ, 825, 88

\bibitem[]{b4}Jose J., Hernanz M., 1998, ApJ, 494, 680

\bibitem[\protect\citeauthoryear{Joshi}{2015}]{b20} Joshi V., Banerjee D. P. K., Ashok N. M., Venkataraman V., Walter F.M., MNRAS, 2015, 452, 3696

\bibitem[]{b5} Joshi, V., Banerjee, D.P.K., 2017, ATel 10369

\bibitem[]{b5} Joshi, V., Banerjee, D.P.K., Srivastava, M., 2017, ATel 10492

\bibitem[]{b5} Munari, u, et al., 2011, MNRAS, 410, L52

\bibitem[]{b43}Pontefract, M., Rawlings, J. M. C. 2004, MNRAS, 347, 1294

\bibitem[Raj et al(2012)Raj et al.]{2012MNRAS...425...2576}
 Raj, A., Ashok, N.M., Banerjee D.P.K., Munari, U.; Valisa, P.; Dallaporta, S., 2012,MNRAS, 425, 2576

\bibitem[]{r1} Raj, Ashish; Banerjee, D. P. K.; Ashok, N. M.; Kim, Sang Chul, 2015, RAA, 15, 993

\bibitem[]{b4}Rawlings J. M. C., 1988, MNRAS, 232, 507


\bibitem[Rudy et al.(2003)Rudy et al.]{2003ApJ...596...1229}
 Rudy R. J., Dimpfl W. L., Lynch D. K., Mazuk S., Venturini C. C., Wilson J. C., Puetter R. C., Perry R. B., 2003, ApJ, 596, 1229

\bibitem[]{r1}Rudy, R. J., Laag, E. A., Crawford, K. B., et al. 2012a, Central Bureau Electronic Telegrams, 3287, 1

\bibitem[]{r1}Rudy, R. J., Russell, R. W., Sitko, M. L., et al. 2012b, Central Bureau Electronic Telegrams, 3103, 1


\bibitem[]{b37} Srivastava, Mudit K.; Ashok, N. M.; Banerjee, D. P. K.; Sand, D., 2015, MNRAS, 454, 1297

\bibitem[]{b37}Srivastava, Mudit K.; Banerjee, D. P. K.; Ashok, N. M.; Venkataraman, V.; Sand, D.; Diamond, T., 2016, MNRAS, 462, 2074

\bibitem[]{b37}Starrfield, S., Truran, J. W., Sparks, W. M., Kutter, G. S. 1972, ApJ, 176, 169

\bibitem[\protect\citeauthoryear{Joshi}{2015}]{b20}Starrfield, S., Gehrz, R. D., Truran, J. W. 1997, in AIP Conf. Proc. 402,
Astrophysical Implications of the Laboratory Study of Presolar Materials,
ed. T. J. Bernatowicz and E. K. Zinner (Woodbury: AIP), 203

\bibitem[]{b4} Strader, J. et al., 2017, Atel 10367

\bibitem[Wichmann et al.(1991)Wichmann et al.]{1991ISRS...353... }
 Wichmann R., Kautter J., Kawara J., Williams R.~E., 1991, in Jaschek C., Andrillat Y., eds, Proc. Int. Coll., Montpellier, France. The Infrared Spectral Region of Stars. Cambridge Univ. Press, Cambridge, p. 353

\bibitem[]{b4} Williams, S.C. , Darnley, M.J., 2017, ATel 10366

\bibitem[]{b4}Wilson O. C., Merrill P. W., 1935, PASP, 47, 53


\end{thebibliography}
\end{document}